\documentclass[printer]{tMOP2e}
\usepackage{psfrag}
\begin{document}
\doi{10.1080/09500340xxxxxxxxxxxx}
 \issn{1362-3044}
\issnp{0950-0340} 

\markboth{J.R. Castrej\'on-Pita et. al.}{J.R. Castrej\'on-Pita et.
al.}

\title{{\itshape Novel Designs for Penning Ion traps}}

\author{J.R. Castrej\'on-Pita, H. Ohadi, D.R. Crick, D.F.A. Winters, D.M. Segal and R.C. Thompson
\newline \smallskip
\thanks{\vspace{6pt}\newline\centerline{\tiny{ {\em Journal of Modern Optics} ISSN 0950-0340 print/ ISSN 1362-3044 online
\textcopyright 2004 Taylor \& Francis Ltd}}
\newline\centerline{\tiny{ http://www.tandf.co.uk/journals}}\newline \centerline{\tiny{DOI:
10.1080/09500340xxxxxxxxxxxx}}} The Blackett Laboratory, Imperial College, Prince Consort Road, SW7 2BZ, London,
United Kingdom.}  \received{$14^{th}$ February 2006}

\maketitle

\begin{abstract}
We present a number of alternative designs for Penning ion traps suitable for quantum information processing (QIP) applications with atomic ions. The
first trap design is a simple array of long straight wires which
allows easy optical access. A prototype of this trap has been built to trap Ca$^+$ and a simple electronic detection scheme has been employed to demonstrate the operation of the trap. Another trap design consists of a conducting plate with a hole in it situated above a continuous conducting plane. The final trap design is based on an array of pad electrodes. Although this trap design lacks the open
geometry of the traps described above, the pad design may prove useful in a hybrid scheme in which information processing and qubit storage take
place in different types of trap. The behaviour of the pad traps
is simulated numerically and techniques for moving ions rapidly
between traps are discussed. Future experiments with these various
designs are discussed. All of the designs lend themselves to the construction of multiple trap arrays, as required for scalable ion trap QIP.
\end{abstract}

\section{Introduction}

A conventional Penning trap consists of three electrodes: a ring
electrode and two endcaps. Ideally these electrodes are
hyperboloids of revolution, producing a quadrupole electric
potential, and traps with hyperbolic electrodes are commonplace.
Trapping of positive ions is achieved by holding the endcaps at a
positive potential with respect to the ring electrode. This
provides one dimensional confinement along the axis of the trap.
The electrode structure is embedded in a strong axial magnetic
field which provides confinement in the other two dimensions
(the radial plane).  This configuration has proven to be useful in
mass-spectrometry measurements and fundamental studies with single
ions \cite{Ghosh}. An important variant of the Penning trap
employs a stack of cylindrical electrodes. Typically five
electrodes are used -- a thin ring electrode, a pair of
`compensation' electrodes (used to trim the potential to achieve a
better approximation to a quadrupole) and a pair of much longer
electrodes that act as the endcaps. The geometry of this arrangement means that it is well suited for use with superconducting solenoid magnets. This design has a
slightly more `open' geometry than the hyperbolic trap since it
has good access along the axis of the trap.  Nonetheless, neither
of these designs are really ideal for applications such as
spectroscopy and QIP studies where optical access is of primary
importance.

A scheme has been proposed that uses a linear array of cylindrical
Penning traps for QIP using trapped electrons \cite{Tombesi}. This
scheme uses microwave techniques rather than optical addressing so
that optical access is not an issue. This approach has limited
scalability since the array of traps is essentially one
dimensional. To address this issue, further work on this scheme is
now being focussed on a new planar trap design \cite{Vogel}.  This
design has an open geometry and is readily scalable to two
dimensional arrays. A trap of this variety has recently been
tested and trapped electrons were successfully detected
\cite{Comm}. The QIP scheme envisaged by Stahl {\it et al.}\cite{Vogel}
involves electrons in different micro-traps being coupled via
superconducting wires.

Another proposal has been made for a very simple trap made using
straight wires \cite{PRA}. This trap shares the optical
accessibility of the Stahl  {\it et al.}\cite{Vogel} design and lends itself,
in a very straightforward way, to the production of arrays of traps.
The basic trap consists of two perpendicular sets of three
parallel straight wires (see figure~\ref{fig:6wire_elec_det}). This trap utilizes an
electrostatic potential between the central and the outer wires to
confine the ions axially, with a magnetic field providing
confinement in the other two dimensions. Another advantage of this
trap is that an analytical expression exists for the equation of
motion of trapped ions. In this paper a first prototype of the
wire trap is presented with experimental evidence of its
operation.

Current ion trap QIP experiments focus on strings of ions in
linear radiofreqency traps, however this approach is unlikely to be
scalable beyond a few tens of ions. The issue of scalability in
ion trap QIP has been carefully addressed in a paper by Kielpinski
\emph{ et al.} \cite{Monroe}. They describe a scheme based on
trapped ions held in multiple miniature rf ion traps. In this
scheme two ions are loaded into a single micro-trap where gate
operations may be performed. Individual ions are then shuttled
into different micro-traps for storage and can be retrieved at a
later time to continue with the calculation. In the last few years
a number of key elements of this approach have been demonstrated
\cite{shiftsplitt}.

Adapting this approach to the Penning trap involves a number of
challenges, however, Penning traps may have some important
advantages. Ambient magnetic field fluctuations have been shown to be the
major limiting cause of decoherence in radiofrequency ion trap QIP to date.
Operating a radiofrequency trap with a small additional magnetic
field at which the `qubit' transition frequency is first-order
B-field independent has been shown to lead to coherence times
greater than 10 seconds \cite{Wineland10sec}. On the other hand,
coherence times of up to 10 minutes have been demonstrated using
similar techniques in a Penning trap \cite{Bollinger10min}.
Another advantage of the Penning trap is that it employs only
static electric and magnetic fields and so heating rates should be
very low in these traps. Shuttling ions between different Penning
traps along the axis of the magnetic field is straightforward
\cite{Werth}, but moving ions in a two dimensional array of traps,
as required for scalability, will inevitably involve moving ions
in a plane perpendicular to the magnetic field. This is
complicated by the presence of the ${\bf v}\times {\bf B}$ term in the Lorentz force.
In this paper we address this issue and describe a possible
architecture for a Penning trap quantum information processor. We
consider two alternative approaches. The first is to allow the ion
to {\em drift} in the desired direction (`adiabatic' approach).
Clearly, speed is a concern for QIP so we have also considered a
`diabatic' approach in which an ion is `hopped' to a desired
location in a single cyclotron loop by the application of a pulsed
nearly linear electric field. In order to be able to switch
between a linear field (for hopping) and a quadrupole potential
(for trapping), a different arrangement of electrodes is required.
The resulting traps are made up from arrays of pad electrodes
whose voltages can be switched rapidly in order to perform the
different required operations.

\section{The wire trap}

\begin{figure}
\centerline{\scalebox{.55}{\epsfbox{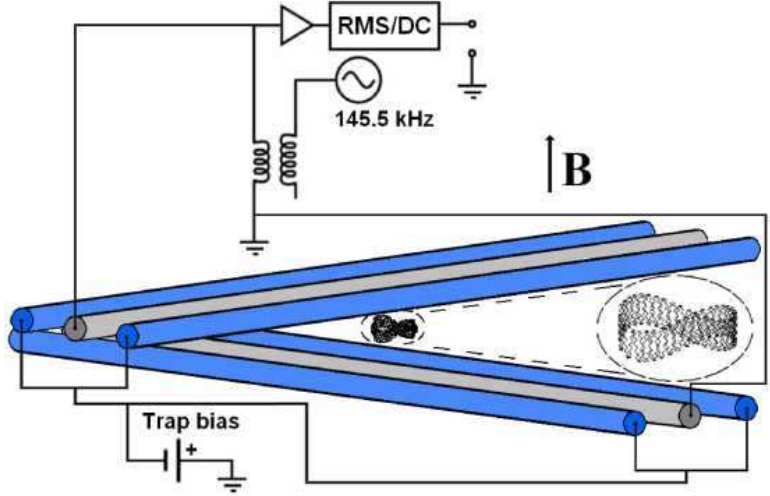}}}\caption{Six wire trap and electronic detection scheme. The simulated trapped ion trajectory shown in
the figure (with an expanded view in the oval inset) corresponds to Ca$^+$ at 10 meV in a trap
with similar dimensions to those of the prototype.  The potential
difference between central and external wires is $\triangle V =
-1.3$ V (at this voltage, the axial frequency of ions corresponds
to the resonant frequency of the detection circuit). The magnetic
field of 1 T is oriented perpendicular to both set of wires.
The simulation  was performed using SIMION. }\label{fig:6wire_elec_det}
\end{figure}

A simple trap based on two sets of three wires or rods is shown schematically in Fig. \ref{fig:6wire_elec_det}. This arrangement of wires produces minima in the electric
potential. At these points, the forces between the charged particle
and the various electrodes (wires) cancel out and thus charged
particles can be three-dimensionally confined with the addition of
a magnetic field in the axial direction. A more complete
description of this trap can be found in \cite{PRA}. Briefly, the
electrostatic potential produced by the six wire ion trap, in
Cartesian coordinates, is

\[
\phi(x,y,z)=-\frac{\triangle V}{2
\ln(d^2/2a^2)}\left(\ln{\frac{R^2}{(x+d)^2+(z+z_0)^2}}-\ln{\frac{R^2}{x^2+(z+z_0)^2}}+\ln{\frac{R^2}{(x-d)^2+(z+z_0)^2}}\right)
\]
\begin{equation}
-\frac{\triangle V}{2 \ln(d^2/2a^2)}\left(
\ln{\frac{R^2}{(y+d)^2+(z-z_0)^2}}-\ln{\frac{R^2}{y^2+(z-z_0)^2}}+\ln{\frac{R^2}{(y-d)^2+(z-z_0)^2}}\right).\label{eq1}
\end{equation}
where $d$ is the distance between two wires, $2z_0$ is the
distance between two sets of wires, $a$ is the diameter of the
wires (a$\ll d$), $R$ is an arbitrary distance at which the
potential is set to zero, $R\gg d$, and $\triangle V$ is the
potential difference between central and external wires
\cite{PRA}. From the latter equation it is possible to find at
least three minima along the axial direction ($z$): above, below
and between the wires. The experimental setup presented here is
focused on the study of trapped ions in the minimum at the centre
of the trap (between the two wire planes), however future work will
investigate the optical detection of ions at all three trapping
points.

\subsection{Experimental setup}
We have built a prototype of this trap and have included a simple electronic detection scheme, also shown in Fig.\ref{fig:6wire_elec_det}. The dimensions of the trap were chosen to fit into standard DN40
CF vacuum components. The `wire' electrodes consist of two sets of
three stainless-steel cylindrical rods of 1 mm diameter and
approximately 35 mm long. Within a
set, the centres of the wires are separated by 3 mm and the centres of the wires are 4 mm apart in
the axial direction. The rods are supported by ceramic mounts and
fixed into an oxygen-free copper base which is mounted on the
electrical feedthrough. The copper base has a 2 mm hole below the
trap center to permit the access of electrons from a filament
placed behind it, but also to shield the trapping volume from the
electric field produced by the filament and its connectors.
Calcium atoms are produced by an atomic beam oven and are ionized by
electron bombardment to produce Ca$^+$. The calcium oven is made of
a 1 mm diameter tantalum tube spot-welded onto a 0.25 mm tantalum
wire. Calcium is placed inside the tantalum tube and then both
ends of the tube are closed. A hole ($\approx$ 0.2 mm diameter) in
the side of the tube allows atoms to effuse into the trapping region when the oven is
heated by an electrical current sent through the tantalum wire.
The electron filament is made of
coiled thoriated tungsten wire with a diameter of 0.25 mm. The trap has
been operated at 2.5 $\times 10 ^{-8}$ mbar and with a magnetic
field of 1 T. The calcium oven is driven at 1.53 Amps and
the filament at 4.4 Amps. The filament is biased by $- 30$ V with respect to the grounded rods to accelerate electrons into the trapping region.

\begin{figure}
\centerline{\scalebox{.50}{\epsfbox{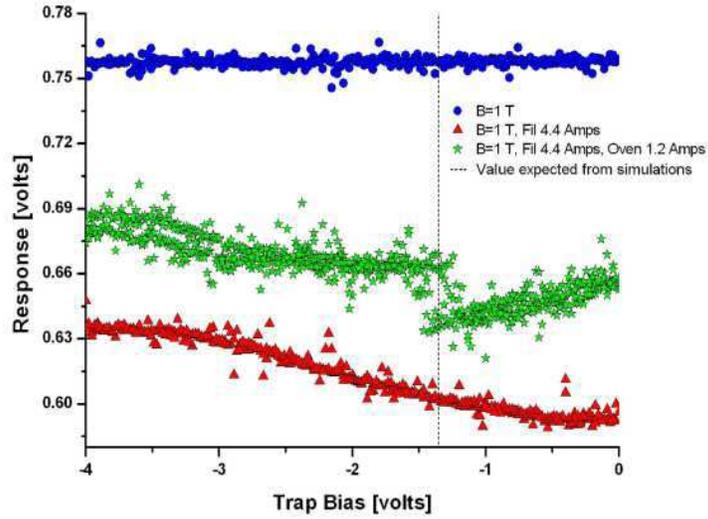}}}\caption{Experimental
results of the electronic detection scheme for
Ca$^+$.}\label{fig:6wire_elec_det_res}
\end{figure}

For the electronic detection of ions, the trap is connected to a
simple resonant circuit where the applied DC trap bias is scanned
so that the axial resonance of the ions in the trap passes through the
resonance of the circuit (see Fig. \ref{fig:6wire_elec_det}). 
The resonant frequency of the circuit is 145.5 kHz and the amplitude of the drive is 10 mV$_{p-p}$. If ions are present
in the trap the quality factor of the resonance is modified as the
ions absorb energy from the circuit. The response of the circuit is monitored and rectified with an rms-DC converter. The presence of ions in the trap should therefore be accompanied by a change in the resulting DC signal. Experimental results are shown in Fig. \ref{fig:6wire_elec_det_res}. The upper curve (circles) is the result when the
electron filament is switched off so that the trapping volume contains no charged particles. The lower curve
(triangles) is the result when the electron filament is on but the
oven, producing calcium atoms, is switched off. The reduction of the signal is due to the presence of electrons in the trapping volume. 
Although this affects the circuit there is no particular resonance since the electrons are not trapped. 
The central curves (stars) result
when both electrons and calcium atoms are present. We note a ubiquitous hysteresis when scanning the voltage from below resonance compared to scanning from above resonance. The resonant feature for calcium ions
is predicted from simulations to occur at -1.35 V and is
observed within the range of $-1.5$ to $-1.1$ V. This initial
result demonstrates that the trap operates successfully and an
experiment to perform laser cooling and optical detection of the
Ca$^+$ ions in this trap is currently being prepared.

\subsection{The two-wire trap}

\begin{figure}
\centerline{\scalebox{.66}{\epsfbox{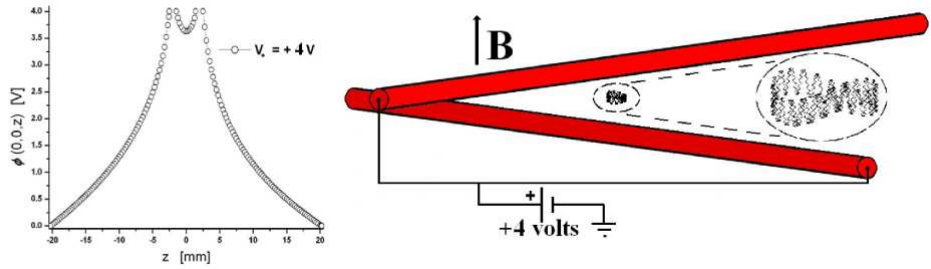}}}\caption{Simulation
of the motion of a $^{40}$Ca$^+$ ion in a two-wire trap performed using SIMION. Ion
kinetic energy 10 meV, $a=0.5$ mm, $z_0=2$ mm and
$B=1$ T. The trajectory is shown enlarged in the oval inset. The potential generated by the trap is shown on the left.} \label{fig:2wire}
\end{figure}

A wire trap can also be formed with just two wires. Using the same notation as in Eq. 1,
the electrostatic potential of two crossed wires separated by
$2z_0$ is
\begin{equation}
\phi =  \frac{V_+}{\ln(R^2/az_0)}\left(\ln
\frac{R^2}{y^2+(z-z_0)^2}+ \ln
\frac{R^2}{x^2+(z+z_0)^2}\right)\label{eq2}
\end{equation}
where $V_+$ is the electric potential of the wires with respect to a ground at $R$. The axial potential in Eq.
\ref{eq2} presents a minimum at $z=0$ where charged particles can
be confined (see Fig. \ref{fig:2wire}).
Following the same procedure as in \cite{PRA}, an approximate
quadratic function around the potential minimum
($x=0$, $y=0$, $z=0$) can be found, giving the result
\begin{equation}
\phi =  \frac{2V_+}{\ln(R^2/az_0)}\left(2 \ln \frac{R}{z_0} -
\frac{1}{z^2_0}\left(x^2+y^2-2z^2\right)\right).\label{eq3}
\end{equation}
If an axial magnetic field ($B$) is added, the equations of motion
of an ion with mass $m$ and charge $q$ around the minimum are then
\[ \ddot{x}= \omega_c\dot{y} + \frac{\omega^2_z}{2} x  \, \, \, , \, \, \,
\ddot{y}= -\omega_c\dot{x}+ \frac{\omega^2_z}{2} y \, \, \, , \,
\, \, \ddot{z}= -\omega^2_z z \] which are the well-known
equations of motion of an ion inside a Penning trap, where
$\omega^2_z = 8qV_+/m z_0^2 \ln (R^2/az_0)$ and $\omega_c = qB/m$.

The two-wire configuration has three disadvantages compared to the
six-wire trap: it does not produce trapping points above
and below the trap, the harmonicity of the potential cannot be modified, and the trap depth is smaller
than for the six-wire trap design. One advantage however is that this trap
would be easier to scale and construct allowing even greater
optical access to the trapped ions. In Fig. \ref{fig:2wire} a
simulation of a trapped Ca$^+$ ion is presented when the electrodes of
a two-wire trap are connected to $+4$ V.

\subsection{`Adiabatic' Ion Transfer in Wire Traps}

\begin{figure}
\centerline{\scalebox{.7}{\epsfbox{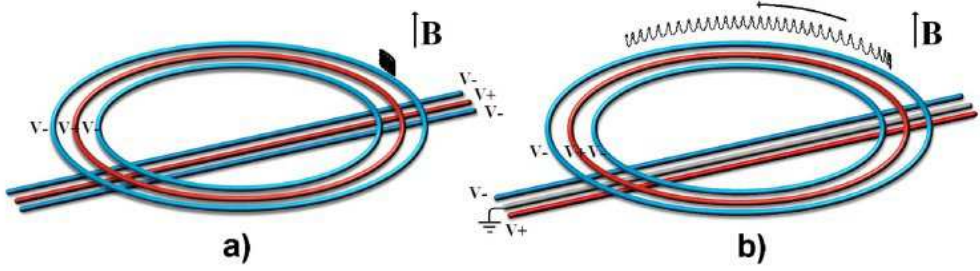}
}}
\caption{Simulations of a trap design to transport ions. In $(a)$
external wires are connected to $-5 V$, and central wires to $+5$
V. In $(b)$ the upper set of wires is connected as in $(a)$, the three lower wires are connected to $-3$ V, 0V, +5 V respectively.  Simulations performed using SIMION for Ca$^+$ with initial kinetic energy 100 meV.} \label{fig:ringtraps}
\end{figure}

Conditions for the transport of ions can be generated using these wire traps. In ref~\cite{PRA} it is shown that a circular `ion guide' can be made using three concentric wire rings with trapping zones along circular lines directly above and below the structure. The goal would be to trap ions at one position and then transport them to another position where they could be
manipulated. One way to achieve this is to use a circular set of three wires as
an ion guide in combination with a straight set of wires, as
illustrated in Fig.~\ref{fig:ringtraps}. In this example, the trapping region
{\em above} the six wire electrodes is used. In Fig. \ref{fig:ringtraps}a
the simulation of a trapped Ca$^+$ ion is observed in the
crossing zone to the right of the figure, where both sets of wires are connected to a potential difference of -10 V. In Fig. \ref{fig:ringtraps}b, the upper set of wires is still connected to the original potential creating an ion
guide along its path, but the lower straight set of wires has been
connected in such a way that they produce a potential gradient
which pushes the ions away from the old trapping region and around
the ring. When the ions reach the left hand crossing region
the voltages on lower set of wires can be adjusted
to the same levels as in Fig. \ref{fig:ringtraps}a producing trapping
conditions again. We term this kind of transport `adiabatic' since the ion undergoes many cyclotron loops as it moves in the electric potential. 

\section{The two-plate trap.}

We also present here a proposal for a single-endcap
trap and computer simulations of it. This proposal is
essentially a modification of a planar trap \cite{Vogel}, but with
a simpler design together with straightforward scalability.

\begin{figure}
\centerline{\scalebox{.6}{\epsfbox{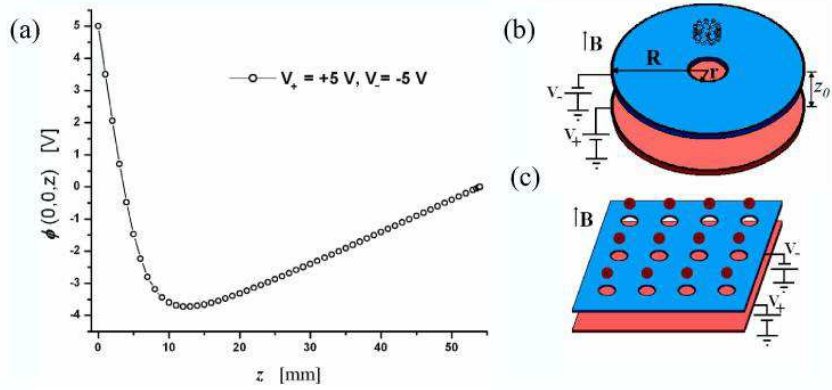}
}}\caption{(a) The axial potential generated by the two-plate trap shown in (b) for $z_0=5$ mm and $R-r=10$mm and with the electrodes connected to $\pm$ 5 volts as shown. The simulated trapped ion trajectory shown in (b) corresponds to a molecular ion with a mass of 100 amu and with an initial kinetic energy of 100 meV in an applied magnetic field of $B=1$ T as shown. The simulation was performed using SIMION. (c) An array of traps based on this design.}
\label{fig:holetrap1}
\end{figure}

The trap presented in \cite{Vogel} consists of a central disk
connected to a positive voltage surrounded by a planar ring
connected to a negative voltage; the entire system is surrounded by a
grounded electrode and is embedded in a planar substrate. 
This design has a number of advantages including the ability to modify the vertical position of the trapping zone individually for each trap in the array, a feature which is essential for the QIP scheme using trapped electrons proposed in ref~\cite{Vogel}. On the other hand making individual connections to an array of traps clearly enormously increases the complexity of the structure. 
In the geometry shown in Fig. \ref{fig:holetrap1}b the two electrodes of the trap described above are deposited on separate planes. An array of such traps can then be generated with only two electrical connections to the entire structure (see figure~\ref{fig:holetrap1}c). Clearly, this has the disadvantage not allowing individual control of the traps but such an array may find uses in any scheme where an array of trapped ions simply acts as a quantum register (for example the `moving head' scheme proposed by Cirac and Zoller \cite{moving_head}.

Essentially this trap is made of two planar electrodes positioned
in different planes ($z=0$ and $z=z_0$). The upper electrode is a
planar ring with outer radius $R$ and inner radius $r$ in the plane $z=z_0$. The lower electrode is a disk of radius $R$ in the plane $z=0$. This trap
can be also understood as a modified Penning trap with one of its
endcaps removed, employing a planar ring electrode and a planar
endcap. This type of configuration is able to produce an axial
trapping potential above the electrodes when they are oppositely
charged; confinement in the radial plane is produced by
the addition of a magnetic field perpendicular to the electrode
planes. To demonstrate this, an axial electrostatic potential is shown in Fig. \ref{fig:holetrap1}a. A simulated trajectory for a trapped ion in this trap are shown in Fig. \ref{fig:holetrap1}b. As mentioned above, this trap geometry, like the two-wire trap described earlier, exhibits great scalability with relative ease of construction and has good optical access due to the open structure.

\section{Pad traps} 

We now consider a different design of trap array that lends itself to the {\em rapid} movement of ions from one trap to another. To see how this may be achieved first consider the motion of an ion in {\em crossed} homogeneous electric and magnetic fields. If we assume the magnetic field is in the positive $z$ direction and a homogeneous electric field is applied in the positive $y$ direction, the trajectory of an ion that starts at rest at the origin is given by the parametric equations
\begin{eqnarray}
x=&-{V_D \over \omega} \sin \omega t + V_Dt\\
y=&{V_D \over \omega}(1-\cos\omega t)
\end{eqnarray}
where $V_D=E/B$ is the `drift velocity' and $\omega=qB/m$ is the cyclotron frequency. The motion is therefore a series of loops in the $xy$ plane such that the ion drifts in the $x$ direction, periodically coming to rest in the $y$ direction. Consider a pair of traps whose axes are along the $z$-direction but whose centres are displaced in the $x$ direction. If the trapping potential could be switched off and replaced by a linear electric field as described above it would be possible to `hop' the ion from the centre of one trap to the centre of the other trap in a single `cycloid loop'. After the `hop' the trapping potential would be re-applied.

An ion completes a single cycloid loop in a time $t=2\pi/\omega$ i.e. in a single cyclotron period. For $^{40}\rm Ca^+$ and $B$=1T we have $t=2.6\mu$s. The size of the cycloid loop is determined by the magnitude of the electric field since the displacement along the $x$ axis as a result of the cycloid loop is given by $x_l=(E/B)t$. For $^{40}\rm Ca^+$ and $B$=1T we have $x_l=2.6\times 10^{-6}E$. Thus an electric field of $\sim$ 1900Vm$^{-1}$ is required for a displacement of 5mm. If the typical trap dimension is of the same order as this displacement, a typical voltage applied to an electrode would be in the region of 20V for a trap with a characteristic dimension of 1cm.

\begin{figure}
\centerline{\scalebox{.42}{\epsfbox{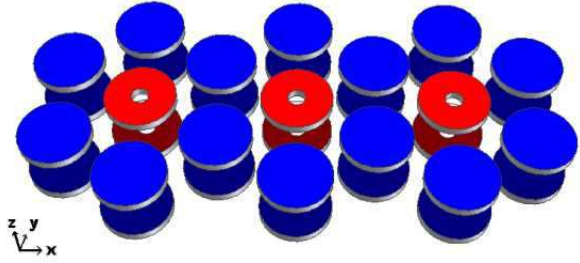}}}
\caption{Schematic layout of  three pad traps. The `endcap' electrodes have holes in them to allow for the loading and extraction of ions in the $z$- direction.}
\label{fig:3pad}
\end{figure}

\begin{figure}
\begin{center}
{\includegraphics[width=12cm]{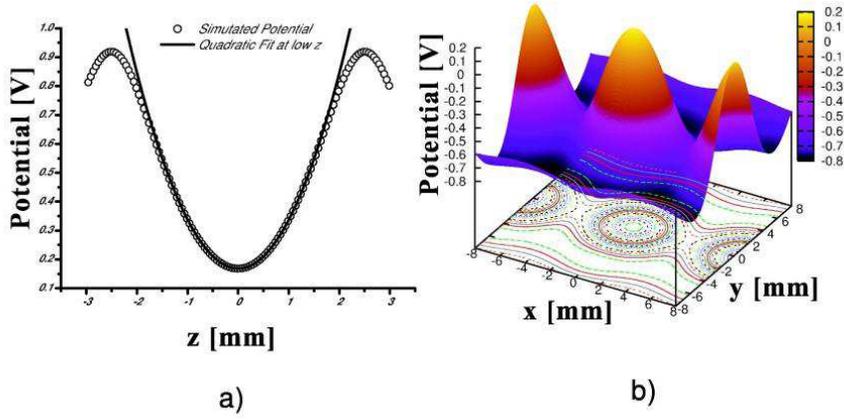}}
\caption{(a) The potential along the axis of the trap for $\gamma=0.9$. The potential only deviates from quadratic behaviour close to the hole in the endcap.  (b) Plot of the potential in the $x,y$ plane at the midpoint between between the substrates. Good circular symmetry is evident for small displacements from the centre of each trap.}
\label{fig:pot_w3}
\end{center}
\end{figure}

The traps must be operable in two distinct modes: a `normal' trapping mode, and a mode in which a nearly linear electric field can be applied across the entire structure (`hopping' mode). The key design criterion is therefore to be able to
switch rapidly between these two modes. The trap array we envisage is made up of two non-conducting substrates onto which a regular pattern of pad electrodes has been deposited. We envisage connections to these electrodes being made through the back of the substrate. We consider a pattern built up from  a `unit cell' that is an equilateral triangular array of three pads. A single quadrupole trap can then be realised in the following way: one substrate would hold a hexagonal array of six pads forming a `ring' with a central pad acting as an `endcap'. The other substrate would carry an identical array of pads. The two opposing layers can then be made to form a trap using 14 electrodes in total -- two endcaps and two sets of six pads making a pair of rings . By applying a positive potential to the endcaps and equal negative potentials to the elements of the rings a quadrupole potential can be generated at the centre of the structure. Other traps can then be made by repeating this pattern in the radial plane (see figure~\ref{fig:3pad}). We have modelled the trapping potential using SIMION. For our simulations we have chosen pads with a diameter of 4mm and a gap between the centres of the pads of 5mm. The potential can be optimised by adjusting the `aspect ratio' of the trap (i.e. the ratio $\gamma$ of the separation of the layers to the separation of the pad centres in the $xy$ plane), such that the deviation of the potential from a pure quadrupole is minimised. Figure~\ref{fig:pot_w3}a shows the deviation of the potential from a pure quadrupole potential for the optimum value of $\gamma=0.9$. The `endcap' electrodes have 1mm central holes to allow for the loading and extraction of ions along the $z$-direction. For the optimum value of $\gamma$ the holes only disrupt the quadrupole potential significantly for relatively large values of $z$. 

Figure~\ref{fig:pot_w3}b shows the potential in the radial plane at the midpoint between the substrates ($z=0$). For a perfect Penning trap the contours projected onto the $x,y$ plane would be circular. The deviation from circular symmetry is small over the trapping region despite the ring being comprised of six separate electrodes.

\begin{figure}
\centerline{\scalebox{.4}{\epsfbox{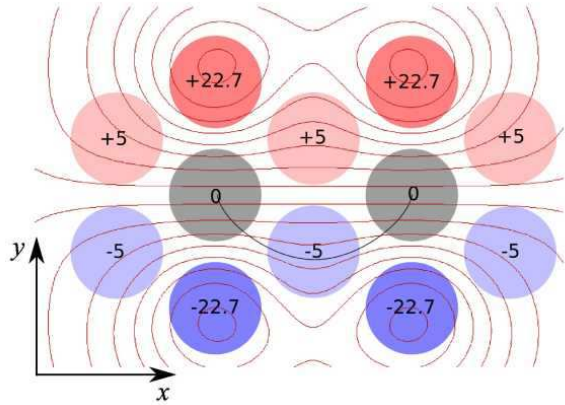}}}
\caption{Applying the potentials shown to the individual pads causes transfer of an ion from the centre of the righthand trap to the centre of the lefthand trap (the magnetic field is along the $+z$ axis i.e. out of the page). The trajectory of an ion initially at rest at the centre of the righthand trap is shown for one cyclotron period.}
\label{fig:2padhopxy}
\end{figure}

In order to `hop' an ion from one trapping site to another, a nearly linear electric field is applied to the trap array. This is done by switching the potentials on the electrodes from the values used for trapping to the ones shown in figure~\ref{fig:2padhopxy}. The figure also shows the contours of the resulting potential in the $z=0$ plane. These potentials are chosen so that a single `cycloid' loop takes a Ca$^+$ ion from one trapping zone to the other assuming a magnetic field of 1T. Since the equipotential surfaces are not exactly planar the resulting trajectory does not have the exact shape of a `cycloid loop', however, provided the ion's velocity has no component in the $y$-direction at the midpoint between the traps, symmetry dictates that the ion will end up in the centre of the second trap. This means that significant deviations from a linear field can be tolerated. In fact, it is possible to choose the potentials applied to the pads in such a way as to generate a more nearly linear electric potential than the one shown in figure~\ref{fig:2padhopxy}.  However, since a genuinely linear potential is not strictly required we have chosen a set of potentials that offer the significant advantage of providing axial ($z$) confinement throughout the `hop'. The trajectory of the ion in the $z=0$ plane is shown in figure~\ref{fig:2padhopxy}. For this trajectory the ion is assumed initially to be at rest at the middle of the rightmost trap.

\begin{figure}
\centerline{\scalebox{.5}{\epsfbox{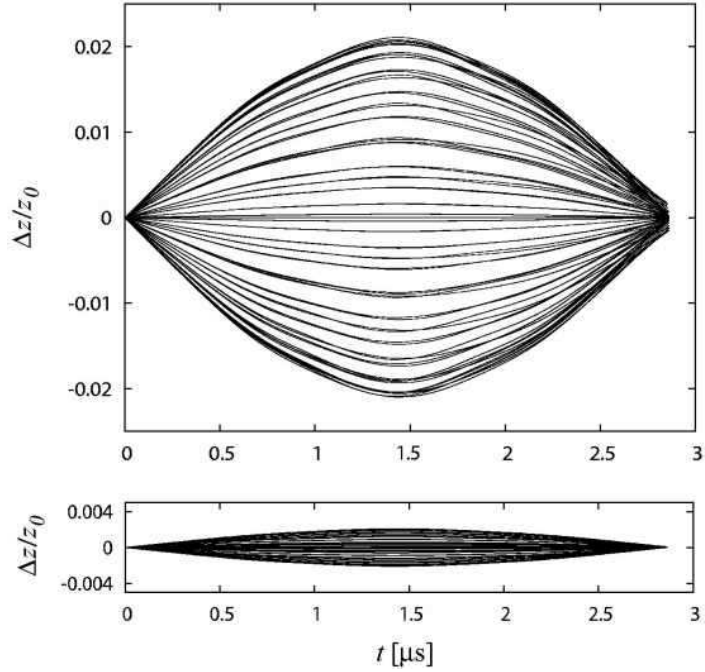}}}
\caption{Simulations (performed using SIMION) of the axial positions of ions $z(t)$ as a function of time, for exactly one cyclotron period. The ions start at the centre of the righthand trap and have a range of initial velocities. For all cases the ions end up near to the centre of the lefthand trap in the $xy$ plane. The upper panel assumes an initial kinetic energy 10 meV. The initial direction of motion is varied over the full range of azimuth and declination in steps of 10 degrees.  The lower panel shows a similar plot but with initial kinetic energy of 0.1 meV. In the $z$-direction the ions are `focused' into the second trap i.e. the final deviation along the $z$-axis  as a result of the hop is small.
}
\label{fig:focus}
\end{figure}

\begin{figure}
\centerline{\scalebox{.5}{\epsfbox{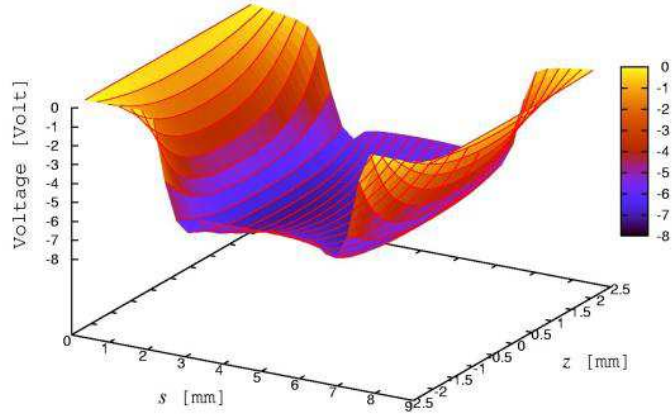}}}
\caption{The potential in the $z$-direction evaluated at a range of equidistant points along the ions trajectory.}
\label{fig:zpot}
\end{figure}

To demonstrate the confinement in the axial direction 
figure ~\ref{fig:focus} shows plots of the axial positions of ions $z(t)$ as a function of time. The ions start at the centre of the   rightmost trap and have a range of initial velocities (see figure caption). Axial confinement is assured because the chosen applied potentials generate a three-dimensional electric potential that has a minimum at $z=0$ all the way along the ion's trajectory in the $x,y$ plane. If we define $s$ as the displacement along this trajectory we can show this by plotting the electric potential along the $z$ axis as a function of $s$. This is shown in figure~\ref{fig:zpot}. The potential is dominated by the steep slopes, first negative and then positive, encountered along the trajectory $s$. This describes the ion first accelerating predominately in the $-y$ direction due to the near-linear electric field, but eventually turning around and climbing back up the electric potential as the ${\bf v}\times{\bf B}$ term becomes significant. On the other hand a close examination of the figure reveals that the potential in the $z$ direction always has a minimum at the position of the ion. The depth of this potential varies along the ion's trajectory so that the $z$- motion is not harmonic during the `hop', but the curvature of the potential ensures axial confinement. Since the ions at all times find themselves in a potential that has a minimum in the $z$ direction they are `focused' into the lefthand trap.

Finally we consider how such an array of `pad' traps could be used for QIP. We envisage a structure in which a stack of `conventional' cylindrical Penning traps could be positioned above one of the pad traps.  Such a linear array of cylindrical traps along the magnetic field direction could be used for gate operations between pairs of ions or multiple ions. Ions could then be ejected from the cylindrical stack into the pad trap directly below it. Ions in the pad trap array could then be shifted sideways allowing for a different ion to be presented to the linear stack of traps.
We intend to build and test a prototype set of pad traps with the dimensions described above. It may be prudent to use closely spaced hexagonal pads rather than circular ones to minimise the exposed are of non-conducting substrate which may be prone to charging up. Ultimately one could use microfabrication techniques to make much smaller pad-traps. The size is only limited by the amount of radial confinement that can be achieved, which depends on the strength of the magnetic field. Furthermore, the geometry of these traps would allow the ions to be axialised\cite{axialisation}, further improving the localisation in the radial plane. Using a 10T superconducting magnet the `hopping' time for Ca$^+$ ions would be 260ns. 
Large arrays of pads could in principle be fabricated, in which case the pads could effectively act as `pixels' allowing almost arbitrary electric fields to be generated. Finally we note that a third option for moving ions in such an array of Penning traps is possible. If rather larger electric fields were applied an ion might be shifted {\em in the direction of the applied electric field} so rapidly that the ${\bf v} \times {\bf B}$ component of the Lorentz force has a negligible bending effect on the trajectories. Small residual sideways shifts of the ion could be compensated by appropriate choice of the direction of the nearly linear electric field applied. Whilst shifts could then in principle be achieved very rapidly, such an approach would require very fine control of the switched potentials, since an ion would need to be first accelerated and finally decelerated as it approached its destination. One of the great advantages of the `hopping' technique considered above is that the ion automatically comes to rest in its new position, and so one can expect rather low heating effects as a result of the shifting of the ion.

\section{Conclusions}

We have presented an experimental setup and experimental evidence for trapped ions inside a simple Penning trap made using a simple array of rods. We have also discussed a variety of other novel Penning traps all of which lend themselves to miniaturisation. We have discussed various strategies for moving ions around in arrays of miniature Penning traps and have set out some of their advantages for applications in QIP. In particular we have described a hybrid quantum information processor based upon a stack of conventional cylindrical Penning traps, used for gate operations, and a two-dimensional array of `pad' traps, in which stored ions act as a quantum register. 

\section{ACKNOWLEDGMENTS}

Project supported by the European Commission within the FP5 RTD
programmes HITRAP and QGATES and the Integrated Project FET/QIPC ``SCALA" FP6. We also acknowledge the support from the EPSRC. JRCP acknowledges the support by CONACyT, SEP and the ORS Awards.

\medskip

\label{lastpage}

\end{document}